\documentclass[12pt, fleqn]{article}
\usepackage[cp1251]{inputenc}
\usepackage{latexsym,amsfonts,amssymb}
\usepackage{graphicx}

\usepackage{amsbsy}
\usepackage{amsmath}
\usepackage{epsf}

\newtheorem{theo}{Theorem}
\newcommand{\bt}{\begin{theo}}
\newcommand{\et}{\end{theo}}
\newcommand{\bd}{\begin{displaymath}}
\newcommand{\ed}{\end{displaymath}}

\newcommand{\be} {\begin{equation}}
\newcommand{\ee} {\end{equation}}
\newcommand{\ba} {\begin{array}}
\newcommand{\ea} {\end{array}}

\newcommand{\p} {\partial}

\newcommand{\lbd} {\lambda}

\begin{document}

\normalsize
 \begin{center}
 {\Large \bf  Lie and conditional symmetries
  of   the  three-component diffusive  Lotka--Volterra system }\\
 \medskip

{\bf Roman Cherniha$^{\dag,\ddag}$} \footnote{\small e-mail:
cherniha@imath.kiev.ua}
 {\bf and  Vasyl' Davydovych$^\dag$}
 \footnote{\small e-mail: davydovych@imath.kiev.ua }
 \\
{\it  $^\dag$~Institute of Mathematics, National Academy
of Sciences of Ukraine,\\
3, Tereshchenkivs'ka Street,  Kyiv 01601, Ukraine}\\
 {\it $^\ddag$~Department of Mathematics, National University `Kyiv Mohyla Academy'\\
2, Skovoroda Street,   Kyiv 04070, Ukraine}\\
\end{center}

\renewcommand{\abstractname}{Abstract}
\begin{abstract}
Lie and  $Q$-conditional symmetries of
 the classical  three-component diffusive   Lotka--Volterra system in the  case of one space
 variable are  studied. The group-classification  problems  for finding  Lie symmetries
 and $Q$-conditional symmetries of the first type are completely  solved.
Notably, non-Lie symmetries ($Q$-conditional symmetry operators)  for   a multi-component non-linear reaction-diffusion system  are constructed   for the first time.
   An example of non-Lie symmetry reduction   for solving
a biologically motivated problem   is  presented.
\end{abstract}


\section{\bf  Introduction}

In 1952  A.C. Turing published the remarkable paper
\cite{turing},  which  a revolutionary idea about mechanism of
morphogenesis (the development of structures in an organism during
the life) has been proposed in. From the mathematical point of view Turing's idea immediately leads to construction of reaction-diffusion systems (not single equations !)  exhibiting so called Turing instability  (see, e.g., Chapter 14.3 in \cite{mur2}).  It should be stressed that  nonlinear reaction-diffusion (RD) systems are govern equations for  many well-known nonlinear  
second-order models  used to describe  
various processes in physics \cite{ames},   
biology \cite{mur2, britton} and ecology \cite{okubo}.
Since 1952
 nonlinear RD
systems
have been extensively studied by means of different mathematical
methods, including group-theoretical methods. Nevertheless finding   Lie symmetries of
two-component RD systems  was initiated about 30 years ago  \cite{zulehner-ames},  complete solving this group-classification problem  was finished  only  a few  years ago  in papers  \cite{ch-king,ch-king2} (for constant diffusivities), \cite{ibrag-94, ch-king4, ch-king06} (for non-constant diffusivities) and \cite{niki-05} (for constant cross-diffusion). It should be stressed that there are only a few   papers devoted
to search for non-Lie (conditional, non-classical) symmetries of  such systems \cite{barannyk2002, ch-pli-08, ch-dav-2011}.
 Because finding non-Lie symmetries for RD systems is very difficult  problem \cite{ch-pli-08, ch-2010}, only  special cases of the  general  two-component RD system were examined in these papers.

All  the  papers cited above, excepting   \cite{ch-king06},
deal with the two-component RD systems. In  section 4  of  \cite{ch-king06}, Lie symmetries of a class of multi-component RD systems
 are completely described.
  To the best of our knowledge there are no any  papers devoted to construction of conditional symmetries for the multi-component RD systems. On the other hand, nonlinear multi-component RD systems is  an important
  tool for  mathematical modeling a wide range of  processes involving several kinds of species (cells, chemicals etc.). Moreover, such systems  possess some properties, which are not common for relevant  two-component systems.
 Thus, it is time to extend  the results obtained for   two-component RD systems on the multi-component systems.
It turns out that this is a highly non-trivial problem and this paper  is devoted to solving this problem for the well-known  three-component RD system.

In the present paper, we shall consider the diffusive Lotka--Volterra
 (DLV) system
 \be\label{2}\ba{l} \lbd_1 u_t =  u_{xx}+u(a_1+b_1u+c_1v+d_1w),\\
\lbd_2 v_t = v_{xx}+ v(a_2+b_2u+c_2v+d_2w),\\ \lbd_3 w_t =
w_{xx}+w(a_3+b_3u+c_3v+d_3w),
\ea\ee  
where  $u(t,x), \ v(t,x), \ w(t,x)$ are unknown concentrations, $a_k, b_k, c_k, d_k$  and $\lambda_k>0$  are arbitrary constants (hereafter $k=1,2,3$ and the  subscripts $t$ and $x$
denote differentiation with respect to these  variable.These constants  have the
relevant biological (chemical) interpretation depending on a type of
interaction between populations (cells, chemicals), which system
(\ref{2}) describes \cite{mur2, britton, okubo}. Setting formally $w=0$, we obtain the two-component
DLV system.   Lie symmetries of
such system  have been completely described in \cite{ch-du-04}, while
our recent paper \cite{ch-dav-2011}  was devoted to search of its
$Q$-conditional symmetries.

System (\ref{2}) is the standard generalization of the classical
Lotka--Volterra  system that takes into account the diffusion process
for interacting species (see terms  $u_{xx}$, $v_{xx}$ and $w_{xx}$).
Nevertheless the classical Lotka--Volterra  system  was independently  introduced by
A. J. Lotka and V. Volterra   about  90 years ago, its different generalizations are  widely studied at the present time because of
their
 importance for  mathematical modeling many  processes in physics, biology, ecology etc.
Notably,  there are several recent papers devoted to rigorous study of  multi-component   diffusion Lotka--Volterra system   (see
 \cite{martinez,jan,wong} and papers cited therein).


\section{\bf Main Definitions }

Because the DLV system  (\ref{2})  belongs to the general class of three-component RD systems
\be\label{1}\ba{l} \lbd_1 u_t =  u_{xx}+C^1(u, v, w),\\
\lbd_2 v_t = v_{xx}+ C^2(u, v, w),\\ \lbd_3 w_t = w_{xx}+C^3(u, v,
w),
\ea\ee
where $C^k(u, v, w)$ are arbitrary smooth functions, we formulate the main definitions for any  system of the form   (\ref{1}).

To find Lie invariance  operators, one needs to consider   system
(\ref{1}) as the manifold ${\cal{M}}=\{S_1=0,S_2=0,S_3=0 \},$  where
\be \nonumber \ba{l}
 S_1 \equiv \ \lbd_1 u_t -  u_{xx}- C^1(u,v,w),\\
S_2 \equiv \ \lbd_2 v_t -  v_{xx}- C^2(u,v,w), \\
S_3 \equiv \ \lbd_3 w_t -  w_{xx}- C^3(u,v,w),\ea\ee in the
prolonged space of the  variables: $t, x, u, v, w, u_t, v_t,w_t$,$
u_{x}, v_{x}, w_x, u_{xx}, v_{xx}, w_{xx}, u_{xt}, v_{xt}, w_{xt}$,
$u_{tt}, v_{tt}, w_{tt}.$ According to the well-known definition, system
(\ref{1}) is invariant under the Lie group generated by the
infinitesimal operator \be\label{2-2}
Q = \xi^0 (t, x, u, v, w)\p_{t} + \xi^1 (t, x, u, v, w)\p_{x} +  
 \eta^1(t, x, u, v, w)\p_{u}+\eta^2(t, x, u, v, w)\p_{v}+\eta^3(t, x, u, v, w)\p_{w},  \ee  
 if the following Lie's invariance conditions are satisfied:
 \be\label{2-3}  
\ba{l}  
\mbox{\raisebox{-1.6ex}{$\stackrel{\displaystyle  
Q}{\scriptstyle 2}$}} (S_1)  
 \equiv  \mbox{\raisebox{-1.6ex}{$\stackrel{\displaystyle  
Q}{\scriptstyle 2}$}}  
\big(\lbd_1 u_t -  u_{xx}- C^1(u,v,w) \big)  
\Big\vert_{\cal{M}}=0, \\[0.3cm]  
\mbox{\raisebox{-1.6ex}{$\stackrel{\displaystyle  
Q}{\scriptstyle 2}$}} (S_2)  
 \equiv  \mbox{\raisebox{-1.6ex}{$\stackrel{\displaystyle  
Q}{\scriptstyle 2}$}}  
\big(\lbd_2 v_t -  v_{xx}- C^2(u,v,w) \big)  
\Big\vert_{\cal{M}}=0, \\[0.3cm]  
\mbox{\raisebox{-1.6ex}{$\stackrel{\displaystyle  
Q}{\scriptstyle 2}$}} (S_3)  
 \equiv  \mbox{\raisebox{-1.6ex}{$\stackrel{\displaystyle  
Q}{\scriptstyle 2}$}}  
\big(\lbd_3 w_t -  w_{xx}- C^3(u,v,w) \big)  
\Big\vert_{\cal{M}}=0,
\ea  
\ee  
where the operator $ \mbox{\raisebox{-1.6ex}{$\stackrel{\displaystyle  
Q}{\scriptstyle 2}$}} $  
is the second  
 prolongation of the operator $Q$ (see, e.g., \cite{ovs, olv,bl-anco-10,fss}).
 \\
 \noindent  \textbf{Definition 1.\cite{ch-2010}}  Operator (\ref{2-2}) is called the
$Q$-conditional symmetry of the first type  for the RD system
(\ref{1}) if  the following invariance conditions are satisfied:
\be\label{2-2*}  
\ba{l}  
\mbox{\raisebox{-1.6ex}{$\stackrel{\displaystyle  
Q}{\scriptstyle 2}$}} (S_1)  
 \equiv  \mbox{\raisebox{-1.6ex}{$\stackrel{\displaystyle  
Q}{\scriptstyle 2}$}}  
\big(\lbd_1 u_t -  u_{xx}- C^1(u,v,w) \big)  
\Big\vert_{{\cal{M}}_1}=0, \\[0.3cm]  
\mbox{\raisebox{-1.6ex}{$\stackrel{\displaystyle  
Q}{\scriptstyle 2}$}} (S_2)  
 \equiv  \mbox{\raisebox{-1.6ex}{$\stackrel{\displaystyle  
Q}{\scriptstyle 2}$}}  
\big(\lbd_2 v_t -  v_{xx}- C^2(u,v,w) \big)  
\Big\vert_{{\cal{M}}_1}=0, \\[0.3cm]  
\mbox{\raisebox{-1.6ex}{$\stackrel{\displaystyle  
Q}{\scriptstyle 2}$}} (S_3)  
 \equiv  \mbox{\raisebox{-1.6ex}{$\stackrel{\displaystyle  
Q}{\scriptstyle 2}$}}  
\big(\lbd_3 w_t -  w_{xx}- C^3(u,v,w) \big)  
\Big\vert_{{\cal{M}}_1}=0,
\ea  
\ee  
where the manifold ${\cal{M}}_1$ is either $\{S_1=0,S_2=0, S_3=0,
Q(u)=0 \}$, $\{S_1=0,S_2=0, S_3=0, Q(v)=0 \}$ or $\{S_1=0,S_2=0,
S_3=0, Q(w)=0 \}$.  Hereafter the notations $\  Q(u) =\xi^0u_t+\xi^1u_x-\eta^1 $, $ Q(v)
= \xi^0v_t+\xi^1v_x-\eta^2$ and $ Q(w) = \xi^0w_t+\xi^1w_x-\eta^3$
are used.\\
\noindent \textbf{Definition 2.\cite{ch-2010}} Operator (\ref{2-2}) is called the
$Q$-conditional symmetry
( non-classical symmetry in terminology used, e.g., in \cite{bl-anco-10})  for the RD system (\ref{1}) if  the
following invariance conditions are satisfied: \be \nonumber
\ba{l}  
\mbox{\raisebox{-1.6ex}{$\stackrel{\displaystyle  
Q}{\scriptstyle 2}$}} (S_1)  
 \equiv  \mbox{\raisebox{-1.6ex}{$\stackrel{\displaystyle  
Q}{\scriptstyle 2}$}}  
\big(\lbd_1 u_t -  u_{xx}- C^1(u,v,w) \big)  
\Big\vert_{{\cal{M}}_3}=0, \\[0.3cm]  
\mbox{\raisebox{-1.6ex}{$\stackrel{\displaystyle  
Q}{\scriptstyle 2}$}} (S_2)  
 \equiv  \mbox{\raisebox{-1.6ex}{$\stackrel{\displaystyle  
Q}{\scriptstyle 2}$}}  
\big(\lbd_2 v_t -  v_{xx}- C^2(u,v,w) \big)  
\Big\vert_{{\cal{M}}_3}=0, \\[0.3cm]  
\mbox{\raisebox{-1.6ex}{$\stackrel{\displaystyle  
Q}{\scriptstyle 2}$}} (S_3)  
 \equiv  \mbox{\raisebox{-1.6ex}{$\stackrel{\displaystyle  
Q}{\scriptstyle 2}$}}  
\big(\lbd_3 w_t -  w_{xx}- C^3(u,v,w) \big)  
\Big\vert_{{\cal{M}}_3}=0,
\ea  
\ee  
where the manifold ${\cal{M}}_3=\{S_1=0,S_2=0, S_3=0, Q(u)=0,
Q(v)=0, Q(w)=0 \}$.

It is easily seen that ${\cal{M}}_3 \subset {\cal{M}}_1 \subset
{\cal{M}}$, hence, each Lie symmetry is automatically a
$Q$-conditional symmetry of the first   type, while each
$Q$-conditional symmetry of the first type is also $Q$-conditional (non-classical) symmetry (see also \cite{ch-2010} for extensive  discussion about a  hierarchy  of conditional symmetry
  operators).


 \textbf{Proposition 1.}  Let us assume that
$X_1=h^2(t,x,u,v,w)\p_v+h^3(t,x,u,v,w)\p_w$ (hereafter $h^2$ and
$h^3$ are the given functions)  is a Lie symmetry operator of the RD
system (\ref{1}) while    $Q_1$ is a $Q$-conditional symmetry of the
first type, which was found using the manifold ${\cal{M}}_1=\{S_1=0, \
S_2=0,\ S_3=0,\ Q(u)=0  \}$.
 Then any linear combination
 $C_1X_1+C_2Q_1$  (hereafter $C_1$ and $C_2$ are arbitrary constants) produces  another  $Q$-conditional symmetry of the first
 type.

 It should be stressed that this statement is not valid  for arbitrary given $Q$-conditional symmetry but only for  that  of the first  type. Proposition 1 will be applied in the next section to prove Theorem 2.


\section{\bf Main Results }

Obviously, the RD system  (\ref{2})  for  arbitrary functions  $C^k(u, v, w)$ admits the two-dimensional Lie algebra,  called the principal (or trivial)   algebra, with the basic operators :
\begin{equation}\label{3-2*} P_t=\p_t, \ P_x=\p_x.
\end{equation}
It can be easily shown that    (\ref{3-2*})  is the principal algebra also for the DLV
system (\ref{2}).
Note that we want to exclude the semi-coupled systems, i.e. those
containing an autonomous equation, hence, hereafter  the
restrictions
\begin{equation}\label{3-3}\ c^2_1+d^2_1\neq0, \ \ b^2_2+d^2_2\neq0, \ \
b^2_3+c^2_3\neq0.\end{equation}
are assumed.

To find all possible extensions of principal  algebra in the case of
the DLV system (\ref{2}), one needs to apply the invariance criteria
(\ref{2-3})  and to solve the obtained system of determining
equations (DEs).
Omitting rather standard calculations we present the  system of DEs in question:
\begin{eqnarray}
&& \xi^0_{x}=\xi^0_{u}=\xi^0_{v}
=\xi^0_{w}=\xi^1_{u}=\xi^1_{v}=\xi^1_{w}=0,
 \label{2-4} \\
 &&\eta^k_{uu}=\eta^k_{uv}=\eta^k_{vv}=\eta^k_{ww}=\eta^k_{uw}=\eta^k_{vw}=0,
\ k=1,2, \label{2-5}\\
&&\eta^1_{xv}=\eta^1_{xw}=\eta^2_{xw}=\eta^3_{xv}=0, \label{2-15} \\
&&(\lambda_1-\lambda_2)\eta^2_u=
(\lambda_1-\lambda_3)\eta^3_u=0,  \label{2-17}\\
&&(\lambda_1-\lambda_2)\eta^1_{v}=
(\lambda_2-\lambda_3)\eta^3_{v}=0,  \label{2-17b}\\
&& (\lambda_1-\lambda_3)\eta^1_{w}=
(\lambda_2-\lambda_3)\eta^2_{w}=0, \label{2-17c} \\
&&2\xi^1_x-\xi^0_t =0,  \label{2-8}\\
&&2\xi^0\eta^2_{xu}+(\lambda_2-
\lambda_1)\xi^1\eta^2_u=0,  \label{2-6} \\
&& 2\xi^0\eta^3_{xu}+(\lambda_3-
\lambda_1)\xi^1\eta^3_u=0,  \label{2-12}\\
&&2\eta^1_{xu}+\lambda_1\xi^1_{t}=0,  \label{2-7}\\
&&2\eta^2_{xv}+\lambda_2\xi^1_t=0,  \label{2-9}\\
&&2\eta^3_{xw}+\lambda_3\xi^1_t=0,  \label{2-13} \\
&&\eta^1C^1_u+\eta^2C^1_v+\eta^3C^1_w+
\eta^1_{xx}-\lambda_1\eta^1_t+(2\xi^1_x-\eta^1_u)C^1-
\eta^1_vC^2-\eta^1_wC^3=0,  \label{2-10}\\
&&\eta^1C^2_u+\eta^2C^2_v+\eta^3C^2_w+
\eta^2_{xx}-\lambda_2\eta^2_t+(2\xi^1_x-\eta^2_v)C^2-
\eta^2_uC^1-\eta^2_wC^3=0,  \label{2-11}\\
&&\eta^1C^3_u+\eta^2C^3_v+\eta^3C^3_w+
\eta^3_{xx}-\lambda_3\eta^3_t+(2\xi^1_x-
\eta^3_w)C^3-\eta^3_uC^1-\eta^3_vC^2=0. \label{2-14}
\end{eqnarray}

Because the functions $C^k(u, v, w)$ have the known structure
defined in  system  (\ref{2})  (otherwise the problem is very
difficult even in the case of two-component RD system
\cite{ch-king,ch-king2}), this system of DEs can be solved in a
straightforward way. In fact,  solving  subsystem
(\ref{2-4})--(\ref{2-15}), one obtains
\be\label{3-2}\ba{l} \xi^0=\xi^0(t), \ \xi^1=\xi^1(t,x), \\
\eta^1=r^1(t,x)u+q^1(t)v+h^1(t)w+p^1(t,x), \\ \eta^2=r^2(t,x)v+q^2(t,x)u+h^2(t)w+p^2(t,x), \\
\eta^3=r^3(t,x)w+q^3(t,x)u+h^3(t)v+p^3(t,x),
\ea\ee  
where $\xi^0, \ \xi^1, \ r^k, \ q^k, \ h^k$  and  $p^k$  are  unknown functions  at the moment.
They can be  found by substitution  (\ref{3-2})  into  (\ref{2-17})--(\ref{2-14}) and  integration of the linear system obtained.
The result  will  depend on the coefficients $a_k, b_k, c_k, d_k$  and $\lambda_k>0$, and can be formulated as theorem 1.

 \bt The  DLV system (\ref{2})
  admits a non-trivial Lie algebra of  symmetries if and only if   one  and the
corresponding  symmetry operator(s)
have the forms listed in table 1. Any other DLV system admitting
three- and higher-order   Lie  algebra is reduced to one of those from table 1 by the
local transformations:
\be\label{3-1}\ba{l} u \rightarrow c_{11}\exp(c_{10}t)u+c_{12}v+c_{13}w,\\
v \rightarrow c_{21}\exp(c_{20}t)v+c_{22}u+c_{23}w,\\ w \rightarrow
c_{31}\exp(c_{30}t)w+c_{32}u+c_{33}v, \\ t \rightarrow
c_{40}t+c_{41}, \ x \rightarrow c_{50}x+c_{51},
\ea\ee  
where  $c_{ij}$  are some correctly-specified constants in each case
($i=1,\dots,5$, $j=0,\dots,3$).
  \et
Let us apply Definition 1 to find  $Q$-conditional symmetries of the
first type of the form (\ref{2-2}) with $\xi^0\neq0$.  In contrary to the standard $Q$-conditional (non-classical) symmetry, here we cannot simply set  $\xi^0=1$ because some operators will be missed as it  was  shown in \cite{ch-2010}. From the formal point of view, one needs to use three
different manifolds in the criteria  (\ref{2-2*}).
 However, we take into account that  the DLV system (\ref{2}) has a symmetric structure and
 admits two discrete transformations of the form $u\rightarrow v, v\rightarrow u, w\rightarrow w$  and
 $u\rightarrow w, v\rightarrow v, w\rightarrow u$. Thus, the algorithm is as follows.

At the first step we apply Definition 1 in the case of the manifold
$\{S_1=0,S_2=0, S_3=0, Q(u)=0 \}$   to find the operators
(\ref{2-2}) with $\xi^0\neq0$. It turns out that the relevant
calculations lead to a system of DEs with the very similar structure to
 (\ref{2-4})--(\ref{2-14}).  There are only two  differences: no equations (\ref{2-17})  and  the equations
 \begin{eqnarray}\nonumber
&&
\eta^1C^2_u+\eta^2C^2_v+\eta^3C^2_w+(\lambda_1-\lambda_2)\frac{\eta^1}{\xi^0}\eta^2_u+
\eta^2_{xx}-\lambda_2\eta^2_t+(2\xi^1_x-\eta^2_v)C^2-\eta^2_uC^1-\eta^2_wC^3=0,
\\ \nonumber
 &&
\eta^1C^3_u+\eta^2C^3_v+\eta^3C^3_w+(\lambda_1-\lambda_3)\frac{\eta^1}{\xi^0}\eta^3_u+
\eta^3_{xx}-\lambda_3\eta^3_t+(2\xi^1_x-\eta^3_w)C^3-
\eta^3_uC^1-\eta^3_vC^2=0 \end{eqnarray} are obtained instead of
(\ref{2-11})--(\ref{2-14}). The simple analysis of these equations
leads to the important conclusion: {\it any  DLV system (\ref{2})
with $\lambda_1=\lambda_2=\lambda_3$
 admits only such   $Q$-conditional symmetries of the
first type, which coincide with  Lie symmetries}. Thus,  we shall consider  system (\ref{2}) with at least two different lambda-s.

The second step of the algorithm consists in solving the obtained
system of DEs. This step can be realized in a similar way as for the
system of DEs  (\ref{2-4})--(\ref{2-14}),
 however, all  solutions leading to the Lie operators should be excluded (it means that  the solutions obtained  should not satisfy equations (\ref{2-17})).
    Moreover, all the   $Q$-conditional symmetries obtained  were reduced to the simplest forms using Proposition 1. Having this done, we arrived at exactly nine DLV systems with correctly-specified coefficients, which admit $Q$-conditional symmetries of the first type.
 The last  step of the algorithm
consists in applying Definition 1 to each of these systems using the
manifold  manifolds $\{S_1=0,S_2=0, S_3=0, Q(v)=0 \}$ and
$\{S_1=0,S_2=0, S_3=0, Q(w)=0 \}$. This allowed to construct all
possible    $Q$-conditional symmetry operators of the first type.
The final result  is  formulated as follows.

  \bt  The DLV system (\ref{2})
   is invariant under $Q$-conditional operator(s)  of the first
type (\ref{2-2}) (with $\xi^0\neq0$) if and only if   one  and the
corresponding operator(s) have the forms listed in table 2. Any
other DLV system admitting a $Q$-conditional operator of the first
type is reduced to one of those from table 2 by a local
transformation from the set  (\ref{3-1}). Simultaneously this
$Q$-conditional operator is transformed to the corresponding
operator listed in table 2 (up to  equivalent representations
generated by   adding  a Lie symmetry operator of the form
$h^2(t,x,u,v,w)\p_v+h^3(t,x,u,v,w)\p_w$).
  \et

\newpage
 {\bf Table 1. Lie symmetry operators of the DLV system  (\ref{2})}
\begin{small}
\begin{center}
\begin{tabular}{|c|c|c|c|
} \hline

  & Reaction terms  &  Restrictions  & Additional Lie symmetries\\
\hline &&&\\ 1 & $u(b_1u+c_1v+d_1w)$&$$&
$D=2t\p_t+x\p_x-2(u\p_u+v\p_v+w\p_w)$
\\
 & $v(b_2u+c_2v+d_2w)$ &$$ &
$$\\
 & $w(b_3u+c_3v+d_3w)$ & $$&\\
\hline &&&\\ 2 & $u(c_1v+d_1w)$&
& $u\p_u$
\\
 & $v(a_2+c_2v+w)$ &$$ &
$$\\
 & $w(a_3+v+d_3w)$ & $$&\\
\hline &&&\\ 3 & $u(c_1v+d_1w)$&
& $u\p_u, \ D$
\\
 & $v(c_2v+w)$ &$$ &
$$\\
 & $w(v+d_3w)$ & $$&\\
\hline &&&\\ 4 & $u(a_1+bu+v)$&$\lambda_2=\lambda_3=1$&
$\exp(-a_2t)v\p_w, \ w\p_w$
\\
 & $v(a_2+u+cv)$ &$$ &
$$\\
 & $w(u+cv)$ & $ $&\\
\hline &&&\\ 5 & $u(bu+v)$&$\lambda_2=\lambda_3=1$& $v\p_w, \
w\p_w, \ D$
\\
 & $v(u+cv)$ &$$ &
$$\\
 & $w(u+cv)$ & $ $&\\
\hline  &&&\\ 6 & $u(a_1+u+v)$&$\lambda_1=\lambda_2=\lambda_3=1,$&
$\exp(-a_1t)u\p_w,$
\\
 & $v(a_2+u+v)$ &$a_1a_2(a_1-a_2)\neq0$&
$w\p_w, \ \exp(-a_2t)v\p_w, $\\
 & $w(u+v)$ & $$&$ (a_2(u+a_1)+a_1v)\p_w$\\
\hline  &&&\\  7 & $u(a+u+v)$&$\lambda_1=\lambda_2=\lambda_3=1,$&
$\exp(-at)u\p_w,$
\\
 & $v(u+v)$ &$a\neq0$&
$w\p_w, \ v\p_w,$\\
 & $w(u+v)$ & $$&$ (u+a+avt)\p_w$\\
\hline &&&\\ 8 & $u(bu+v)$&$\lambda_1=\lambda_2=\lambda_3=1,$& $D, \
w\p_w, \ \big((b-1)u+(1-c)v\big)\p_w$
\\
 & $v(u+cv)$ &$b\neq1, \ c\neq1 $&
$$\\
 & $w(bu+cv)$ & $$&$$\\
\hline

\end{tabular}
\end{center}
\end{small}

\bigskip

\newpage

{\bf Table 2. $Q$-conditional symmetries  of the first type
of the DLV system (\ref{2})}
\begin{small}
\begin{center}
\begin{tabular}{|c|c|c|c|
} \hline

  & Reaction terms  &  Restrictions   &$Q$-conditional symmetry operators  \\

\hline &&&\\ 1 & $u(a_1+bu+bv+dw)$&$a_1\neq a_2,$&
$\p_t+\frac{a_1-a_2}{\lambda_1-\lambda_2}u(\p_u-\p_v), $
\\
 & $v(a_2+bu+bv+dw)$ &$(b-1)^2+(d-d_3)^2\neq0$ &
$\qquad \p_t+\frac{a_1-a_2}{\lambda_1-\lambda_2}v(\p_v-\p_u)$\\
 & $w(a_3+u+v+d_3w)$ & $$&$$ \\
\hline &&&\\ 2 & $u(a_1+u+v+w)$&$(a_1-a_2)^2+(a_1-a_3)^2\neq0$&
$Q^2_i, \ i=1, \dots , 6$
\\
 & $v(a_2+u+v+w)$ &$$ &
$$\\
 & $w(a_3+u+v+w)$ & $$&\\
\hline &&&\\ 3 &
$u(a_1+u+v+w)$&$(\lambda_2-\lambda_3)a_1-\lambda_2a_3+\lambda_3a_2=0,$&
$Q^2_i, \ i=1, \dots , 6,$
\\
 & $v(a_2+u+v+w)$ &$a_2\neq a_3,$ &
$\p_t+\beta\exp\Big(\frac{a_2-a_3}{\lambda_2-\lambda_3}t\Big)u(\p_v-\p_w)$\\
 & $w(a_3+u+v+w)$ & $\beta\neq0 $&\\
\hline &&&\\4 &
$u(a_1+u+v+w)$&$(\lambda_2-\lambda_3)a_1-(\lambda_1-\lambda_3)a_2+$&
$Q^4_i, \ i=1, \dots , 6$
\\
 & $v(a_2+u+v+w)$ &$(\lambda_1-\lambda_2)a_3=0,$ &
$$\\
 & $w(a_3+u+v+w)$ & $(a_1-a_2)^2+\alpha^2\neq0 $&$$\\
\hline &&&\\5 & $u(a_1+bu+v)$&$(b-1)^2+(c-1)^2\neq0$& $Q^5_1$
\\
 & $v(a_2+u+cv)$ &$$ &
$$\\
 & $w(bu+v)$ & $$&\\
\hline &&&\\ 6 & $u(a_1+u+v)$&$$& $Q^5_1, \ Q^6_i, \ i=1, \dots , 4$
\\
 & $v(a_2+u+v)$ &$$ &
$$\\
 & $w(u+v)$ & $ $& $$\\
\hline &&&\\ 7 & $u(a_1+bu+cv)$&$ \lambda_2=\lambda_3=1,$&
$\p_t+\big((1-b)u+(1-c)v+a_2(1-c)\big)\p_w$
\\
 & $v(a_2+u+v)$ &$b\neq1, \ c\neq1,$ &
$$\\
 & $w(bu+v)$ & $ a_1(1-b)=a_2b(1-c)$&\\
\hline &&&\\ 8 & $u(a+bu+cv)$&$\lambda_2=\lambda_3=1,$&
$\p_t+(1-c)\p_w+$
\\
 & $v(a+u+v)$ &$b\neq1, \ c\neq1,$ &
$\big((1-b)u+(1-c)v\big)\varphi_4(t)\p_w$\\
 & $w(bu+v)$ & $b(2-c)=1$&$$\\
\hline &&&\\  9 & $u(a_1+u+v)$&$\lambda_2=\lambda_3=1$& $Q^9_i, \
i=1, \dots , 5$
\\
 & $v(a_2+u+v)$ &$$&
$$\\
 & $w(u+v)$ & $$&$$\\
\hline

\end{tabular}
\end{center}
\end{small}

\newpage

In table 2, the following designations are introduced:
\[Q^2_i=Q^4_i \ $with$ \  \alpha=0, \ i=1, \dots, 6;\]
\[ Q^4_1=\p_t+\frac{a_1-a_2}{\lambda_1-\lambda_2}u(\p_u-\p_v)+\alpha u(\p_v-\p_w), \
Q^4_2=\p_t+\frac{a_1-a_2}{\lambda_1-\lambda_2}v(\p_v-\p_u)+\alpha
v(\p_u-\p_w),\]
\[Q^4_3=\p_t+\frac{a_1-a_3}{\lambda_1-\lambda_3}u(\p_u-\p_w)+\alpha
u(\p_v-\p_w), \
Q^4_4=\p_t+\frac{a_1-a_3}{\lambda_1-\lambda_3}w(\p_w-\p_u)+\alpha
w(\p_u-\p_v)
 , \]
\[Q^4_5=\p_t+\frac{a_2-a_3}{\lambda_2-\lambda_3}v(\p_v-\p_w)+\alpha
v(\p_u-\p_w), \
Q^4_6=\p_t+\frac{a_2-a_3}{\lambda_2-\lambda_3}w(\p_w-\p_v)+\alpha
w(\p_u-\p_v) ;\]
\[Q^5_1=\p_t+\alpha_1\p_x+ \exp\Big((\frac{(\lambda_1-\lambda_3)^2}{4}\alpha^2_1-
a_1)\frac{t}{\lambda_3}+\frac{\lambda_1-\lambda_3}{2}\alpha_1x\Big)u\p_w;
 \]
 \[Q^6_1=\p_t+\alpha_1\p_x+\exp\Big((\frac{(\lambda_2-\lambda_3)^2}{4}\alpha^2_1-a_2)\frac{t}{\lambda_3}+
\frac{\lambda_2-\lambda_3}{2}\alpha_1x\Big)v\p_w,\]
 \[Q^6_2=\p_t+\frac{a_1-a_2}{\lambda_1-\lambda_2}u(\p_u-\p_v)+
 \beta\exp\Big(\frac{(\lambda_1-\lambda_3)a_2-(\lambda_2-\lambda_3)a_1}{\lambda_3(\lambda_2-\lambda_1)}t\Big)u\p_w,\]
\[Q^6_3=\p_t+\frac{a_1-a_2}{\lambda_1-\lambda_2}v(\p_v-\p_u)+
\beta\exp\Big(\frac{(\lambda_2-\lambda_3)a_1-(\lambda_1-\lambda_3)a_2}{\lambda_3(\lambda_1-\lambda_2)}t\Big)v\p_w,\]
\[Q^6_4=\p_t+\frac{a_2\lambda_1-a_1\lambda_2}{\lambda_3(\lambda_2-\lambda_1)}w\p_w+
\exp\Big(\frac{(\lambda_3-\lambda_2)a_1-(\lambda_3-\lambda_1)a_2}{\lambda_3(\lambda_1-\lambda_2)}t\Big)w(\p_u-\p_v);\]
\[ Q^9_1=Q^5_1 \ $with$ \ \lambda_3=1,
\quad Q^9_2=Q^6_4 \ $with$ \ \lambda_2=\lambda_3=1,\]
\[Q^9_{3}=\p_t+\frac{a_1-a_2}{\lambda_1-1}u(\p_u-\p_v)+(\varphi_1(t)u+\varphi_2(t)v+\beta_1)\p_w, \]
\[Q^9_{4}=\p_t+(\varphi_3(t)u+\varphi_2(t)v+\beta_1)\p_w,
\quad Q^9_5=\p_t+\frac{a_1-a_2}{\lambda_1-1}v(\p_v-\p_u);\]
 where the functions $\varphi_i(t) \
(i=1,\dots,4)$:
\[ \varphi_1(t)=
\begin{cases} \beta_1t + \beta_2,  & $if$ \ a_2=0, \\  \beta_2\exp(-a_2t)+\frac{\beta_1}{a_2},
  & $if$ \  a_2\neq0;  \end{cases}
\qquad \varphi_2(t)=
\begin{cases} \beta_1t,  & $if$ \ a_2=0, \\ \frac{\beta_1}{a_2},
  & $if$ \  a_2\neq0;  \end{cases} \]
\[ \varphi_3(t)=
\begin{cases} \beta_1t + \beta_2,  & $if$ \ a_1=0, \\  \beta_2\exp(-a_1t)+\frac{\beta_1}{a_1},
  & $if$ \  a_1\neq0;  \end{cases}
\qquad \varphi_4(t)=
\begin{cases}t + \beta, & $if$ \ a=0, \\  \beta\exp(-at)+\frac{1}{a},
  & $if$ \  a\neq0  \end{cases} \]
  while  $ \alpha$ and $\beta$ (with and without subscripts 1 and 2 )  are arbitrary constants.

  Finally, we point out for the reader that  {\it (i)} the inequalities listed  in the second  column of table 2 guarantee  that the relevant  operators from the third column are not equivalent to any  Lie  symmetry operators listed in table 1; {\it (ii)} if the given  DLV system from  table 2 contains as a particular case (up to
local transformations of the form (\ref{3-1})) a simpler  system listed in another case of  table 2 then, in order to get the complete list of symmetries, one should go the case with the simpler system .

\textbf{Example.} Let us apply the results obtained to a biologically motivated system of the form  (\ref{2}). One notes that the  DLV system in case 4 of table 2 is equivalent (the relevant substitution is $u \to -bu, \ v \to -cv, \ w \to -dw$)   to  the system
 \be\label{3}\ba{l} \lbd_1 u_t =  u_{xx}+u(a_1-bu-cv-dw),\\
\lbd_2 v_t = v_{xx}+ v(a_2-bu-cv-dw),\\ \lbd_3 w_t =
w_{xx}+w(a_3-bu-cv-dw),
\ea\ee  
where the coefficients $a_k, \ b, \ c$ and $d$ are the known positive constants. It is well-known that system (\ref{3}) is used  for modeling  competition between three species in population dynamics \cite{mur2,britton}.  Substituting $u \to -bu, \ v \to -cv, \ w \to -dw$  into    the $Q$-conditional symmetry operator $Q^4_1$ (see case 4 of table 2)  one obtains  \be\label{4}Q^4_1  \to
Q=\p_t+\frac{a_1-a_2}{\lambda_1-\lambda_2}\,u(\p_u-\frac{b}{c}\,\p_v)+\alpha b\,
u\big(\frac{1}{c}\,\p_v-\frac{1}{d}\,\p_w\big). \ee
 Applying the standard procedure for reducing the given PDE system  to a  ODE system  via  the known symmetry  operator  (\ref{4}), we easily find
 the ansatz
\be\label{7}\ba{l} bu=\varphi_1(x)\exp\big(\frac{a_1-a_2}{\lambda_1-\lambda_2}t\big), \medskip \\
cv=\varphi_2(x)+
(\alpha\frac{\lambda_1-\lambda_2}{a_1-a_2}-1)\varphi_1(x)\exp\big(\frac{a_1-a_2}{\lambda_1-\lambda_2}t\big), \medskip \\
dw=\varphi_3(x)-\alpha\frac{\lambda_1-\lambda_2}{a_1-a_2}\varphi_1(x)\exp\big(\frac{a_1-a_2}{\lambda_1-\lambda_2}t\big), \, a_1\neq a_2\ea\ee
where  $\varphi_1(x)$, $\varphi_2(x)$ and  $\varphi_3(x)$  are new
unknown functions. Substituting  ansatz (\ref{7}) into (\ref{3}) and taking into account the restriction $(\lambda_2-\lambda_3)a_1-(\lambda_1-\lambda_3)a_2+(\lambda_1-\lambda_2)a_3=0$ (see case 4 of table 2),
one obtains reduced system of ODEs
\be\label{8}\ba{l}\varphi_1''+\varphi_1\big(\frac{\lambda_1a_2-\lambda_2a_1}{\lambda_1-\lambda_2}-\varphi_2-\varphi_3\big)=0, \medskip \\
\varphi_2''+\varphi_2\big(a_2-\varphi_2-\varphi_3\big)=0, \medskip\\
\varphi_3''+\varphi_3\big(a_3-\varphi_2-\varphi_3\big)=0.\ea\ee
Now   exact solutions of the 3-D competition system (\ref{3}) can be easily derived by inserting   solutions of this  system into ansatz (\ref{7}).
For example, the constant  solution $\varphi_2=v_0, \ \varphi_3=a_2-v_0$ of the second and third equations of   (\ref{8}) with $a_2=a_3$, produces the following exact solution of the  3-D competition system (\ref{3}) with $ a_1\neq a_2=a_3$:
\be\label{7a}\ba{l} u=\frac{1}{b}\varphi_1(x)\exp\big(\frac{a_1-a_2}{\lambda_1-\lambda_2}t\big), \medskip \\
v=\frac{v_0}{c}+ \frac{1}{c}
(\alpha\frac{\lambda_1-\lambda_2}{a_1-a_2}-1)\varphi_1(x)\exp\big(\frac{a_1-a_2}{\lambda_1-\lambda_2}t\big), \medskip \\
w=\frac{a_2-v_0}{d} -\frac{\alpha}{d}\frac{\lambda_1-\lambda_2}{a_1-a_2}\varphi_1(x)\exp\big(\frac{a_1-a_2}{\lambda_1-\lambda_2}t\big), \ea\ee
where $\varphi_1(x)$ is a solution of the linear ODE $ \varphi_1''+\frac{\lambda_2(a_2-a_1)}{\lambda_1-\lambda_2}\,\varphi_1=0. $
It should be stressed that this solution is not obtainable by any Lie symmetry because system (\ref{3})
  is invariant only under  the principal algebra (\ref{3-2*}), so that only traveling wave solutions can be constructed.

\section{\bf  Conclusions }

In this communication  Lie and   $Q$-conditional symmetries for
 the 3-D diffusive Lotka--Volterra   system  (\ref {2})
 are studied. The main results  are  presented in theorems 1 and 2 giving the
exhaustive list of the systems admitting   Lie  symmetry  and $Q$-conditional symmetry of the first type, respectively.
To the best of our knowledge,  there is only paper \cite{nucci96} (see section 4.1)  where the authors  attempted to construct   $Q$-conditional symmetry operators for a three-component PDE system, the classical  Prandtl  system, using so called the iterating non-classical method.  However, all the results obtained  therein are obtainable  by the  Lie method (this is clearly indicated in  \cite{nucci96}).
Here the $Q$-conditional symmetry operators listed in table 2 are not reducible to Lie operators from table 2.  {\it Thus,  we have constructed non-Lie symmetries for   a multi-component non-linear system  of PDEs  for the first time}. Moreover,  we have applied the $Q$-conditional symmetry operator (\ref{4}) to reduce the relevant Lotka--Volterra   system  and shown that  even  the simplest particular solution of the  ODE system obtained leads to  the exact solution  (\ref {7}), which cannot be derived by  any Lie symmetry.



\end{document}